\documentclass[a4paper,12pt]{article}
\usepackage[cp1251]{inputenc}
\usepackage[russian]{babel}

\newcommand{\be}{\begin{equation}}
\newcommand{\ee}{\end{equation}}
\begin{document}

{\Large\bf Detection of dark energy near the Local Group}

\hspace{2.4cm} {\Large\bf  with the {\it Hubble Space Telescope}}

\vspace{1cm}

{ A.D.~Chernin$^{1,2}$, I.D.~Karachentsev$^3$, P.~Teerikorpi$^2$,
M.J.~Valtonen$^{2}$, G.G.~Byrd$^4$, Yu.N.~Efremov$^1$ ,
V.P.~Dolgachev$^1$, L.M.~Domozhilova$^1$, D.I.~Makarov$^3$,
Yu.V.~Baryshev$^5$}

\vspace{1cm} {\it $^1$Sternberg Astronomical Institute, Moscow
University,Moscow, 119899, Russia,

$^2$Tuorla Observatory, Turku University, Piikki\"o, 21 500,
Finland,

$^3$Special Astrophysical Observatory, Nizhnii Arkhys, 369167,
Russia,

$^4$University of Alabama, Tuscaloosa, USA,

$^5$Astronomical Institute, St.Petersburg University, 198504,
Russia}

\vspace{1cm}

{\bf \noindent We report the detection of dark energy near the
Milky Way made with precision observations of the local Hubble
flow of expansion. We estimate the local density of dark energy
and find that it is near, if not exactly equal to, the global dark
energy density. The result is independent of, compatible with, and
complementary to the horizon-scale observations in which dark
energy was first discovered. Together with the cosmological
concordance data, our result forms direct observational evidence
for the Einstein antigravity as a universal phenomenon -- in the
same sense as the Newtonian universal gravity.}

\vspace{1cm}

Dark energy is the mysterious form of cosmic energy that produces
antigravity and accelerates the global expansion of the universe.
It was first discovered ({\it 1,2}) in 1998-99 in observations of
the Hubble expansion flow with the use of type Ia supernovae at
horizon-size distances of more than 1000 megaparsec (Mpc)  (1 Mpc
is equal to 3.26 million light-years). These and other studies,
especially the observations of the cosmic microwave background
(CMB) anisotropy ({\it 3,4}), indicate that the global dark energy
density is $(0.75 \pm 0.05)\times 10^{-26}$ kilograms per cubic
meter (kg/m$^3$). It contributes nearly 3/4 the total energy of
the universe ({\it 1-4}). According to the simplest,
straightforward and quite likely interpretation, dark energy is
described by the Einstein cosmological constant. If this is so,
dark energy is the energy of the cosmic vacuum ({\it 5}) with the
equation of state $p_V = - \rho_V$. Here $\rho_V, p_V$ are the
dark energy density and pressure which are both constant in time
and uniform in space (the speed of light $c = 1$ hereafter). The
interpretation implies that although dark energy betrayed it
existence through its effect on the universe as a whole, it exists
everywhere in space with the same density and pressure. How to
examine this in direct observations on smaller spatial scales?

We have searched for dark energy in our closest galactic
neighborhood. The local space volume is dominated by our Milky Way
and its sister galaxy, M31, located at about 0.7 Mpc from us,
moving toward each other with a relative velocity $\sim$ 100 km/s.
Together with the Magellanic Clouds, the Triangulum galaxy and
about four dozen other dwarf galaxies, these two major galaxies
form the Local Group. Around the group,two dozen dwarf galaxies are
seen which all move apart of the group. This is the local
expansion flow discovered in the late 1920s by Hubble.

Systematic observations of distances and motions of galaxies in
the Local Group and in the flow around it have been carried out
over the last eight years with the {\it Hubble Space Telescope}
during more than 200 orbital periods({\it 6-12}). High precision
measurements were made of the radial velocities (with 1-2 km/s
accuracy) and distances (8-10 \% accuracy) for about 200 galaxies
of the Local Group and neighbors from 0 to 7 Mpc from the group
barycenter.

We have focused on the shortest distances less than 3 Mpc from the
Local Group barycenter. This is the very beginning of the Hubble
flow of expansion. The flow is represented in Fig.1 by the plot of
radial velocities versus distance, and this is the most complete
version of the Hubble diagram for these scales up to date. The
velocities and distances are given in the reference frame of the
barycenter of the Local Group. At less than 1 Mpc, one sees the
internal, gravitationally-dominated motions of galaxies within the
group. Most of the galaxies are gathered in two families around
the major members of the group. The total mass of the group is
estimated as $M = 1.3 \pm 0.3\times 10^{12} M_{\odot}$ ({\it 10}).

It is seen from Fig.1 that the expansion flow takes over at a
distance $\simeq$ 1 Mpc, just at the outskirts of the Local Group.
A linear velocity-distance trend, $V \propto R$, known as the
Hubble law, emerges at about 2 Mpc distance. The measured value of
the local expansion rate (the Hubble parameter) is $H_0 = 72 \pm
6$ km/s/Mpc ({\it 11}). The flow is rather regular and ``cool'':
its radial one-dimensional velocity dispersion is remarkably low,
17 km/s ({\it 9}).

Like in the largest-scale studies ({\it 1,2}), we use the observed
expansion flow as a natural tool for probing dark energy. The
dwarf galaxies of the flow are good ``test particles'' which may
reveal for us the dynamics behind the observed flow motion. Each
particle is affected by the gravitational attraction of the Local
Group. Considering only the most important dynamical factors, we
may take the gravity field of the group as nearly
centrally-symmetric and static; this is a good approximation to
reality, as exact computer simulations prove ({\it 13,14}).
According to Newtonian gravity law, this force gives a particle
acceleration (force per unit mass)
% 1
\be F_N = - GM/R^2,\ee at its distance $R$ from the group
barycenter.

We consider a picture in which the Local Group and the expansion
flow around it are all embedded in the dark energy with a uniform
local density $\bar\rho_V$ which is, generally, not necessarily
equal to the global density $\rho_V$. Respectively, each particle
of the flow is also affected by the repulsive antigravity force
produced by the local dark energy background. This force can be
described in terms of Newtonian mechanics as well, and according
to the `Einstein antigravity law', the dark energy gives
acceleration
% 2
\be F_E  = G 2 \bar\rho_V (\frac{4 \pi}{3}R^3)/R^2 =  \frac{8
\pi}{3} G\rho_V R, \ee \noindent where $-2 \bar\rho_V = \bar\rho_V
+ 3 \bar p_V$ is the local effective (General Relativity)
gravitating density of dark energy (for details see ({\it 15} )
where a General Relativity treatment is also given). The local
pressure of dark energy is negative, $\bar p_V$, and so the
effective gravitating density is negative as well. Because of this
the acceleration is positive, and it speeds up the particle motion
apart from the center.

It is seen from Eqs.1 and 2 that the gravity force ($\propto
1/R^2$) dominates over the antigravity force ($\propto R$) at
small distances, and here the total acceleration is negative. At
large distances, antigravity dominates, and the acceleration is
positive there.  Gravity and antigravity balance each other, and
so the acceleration is zero, at the ``zero-gravity surface'' which
has a radius
% 3
\be R_V =  (\frac{3 M}{8\pi \bar\rho_V})^{1/3}. \ee

If one takes into account the real structure of the Local Group,
it may be seen ({\it 13,14}) that the zero-gravity surface is not
exactly spherical and not exactly static; but it is nearly
spherical and remains almost unchanged (within the 15-20\%
accuracy) since the formation of the Local group some 12 Gyr ago,
as the computer simulations indicate.

The model described by Eqs.1-3 is obviously very different from
the Friedmann cosmological model of a uniform and isotropic
universe. And this must be so, because there is no uniformity or
isotropy on the spatial scale of a few Mpc. Moreover, the force
field of the universe as a whole is non-stationary and changing
with time, while the local force field (given by Eqs.1-2) is
static. Consequently, the motion of the local flow galaxies hardly
originated in the global initial isotropic Big Bang; its nature is
rather essentially local and caused by the local processes. One
may imagine that the flow galaxies gained their initial velocities
in the early days of the Local Group when its major and minor
galaxies participated in violent non-linear dynamics with multiple
collisions and mergers. In this process, some of dwarf galaxies
managed to escape from the gravitational pool of the Local Group
after having gained escape velocity from the non-stationary
gravity field of the forming group. This process is suggested by
the concept of the ``Little Bang''({\it 16}) and supported by the
computer simulations ({\it 13,14}).

When escaped particles occur beyond the zero-gravity surface ($R
> R_V$), their motion is controlled mainly by the dark energy
antigravity. The general trend of the dynamical evolution of the
flow may be seen from Eqs.1-3. At large enough distances where
antigravity dominates over gravity almost completely, the
velocities of the flow are accelerated and finally they grow with
time exponentially: $V \propto \exp [H_V t]$. At this limit, the
distances grow exponentially as well. As a result, the expansion
flow acquires the linear velocity-distance relation
asymptotically: $V \rightarrow H_V R$. Here the value
% 4
\be H_V = (\frac{8 \pi G}{3} \bar\rho_V)^{1/2} \ee \noindent is
the expansion rate which is constant and determined by the local
dark energy density alone.

The zero-gravity radius $R_V$ is obviously the key physical
quantity in this picture. How to find its value in the observed
expansion flow? Basing on the dynamics considerations above, we
may robustly restrict the value of $R_V$ with the use of the
diagram of Fig.1. Indeed, since the zero-gravity surface lies
outside the Local Group volume, it should be that $R_V > 1 $ Mpc.
On the other hand, the fact that the linear velocity-distance
relation is seen from a distance of about 2 Mpc suggests that $R_V
< 2$ Mpc. If so, Eq.3 leads directly to the robust upper (from
$R > 1$ Mpc) and lower ($R < 2$ Mpc) limits to the local dark
energy density:
% 5
\be (0.1 \pm 0.03) < \bar\rho_V < (1 \pm 0.3) \times 10^{-26} \;\;
kg/m^3. \ee \noindent (Here the measured value of the Local Group
mass is also used.)

The lower limit in Eq.5 is most significant. It means that the
dark energy does exist in the nearby universe. In combination,
both limits imply that the value of the local dark energy density
is near the value of the global dark energy density, $\bar\rho_V
\sim \rho_V$, or may be exactly equal to it. Anyway, the global
figure for $\rho_V$ ($(0.75 \pm 0.05) \times 10^{-26}$ kg/m$^3$ --
see above) lies comfortably in the range given by Eq.5.

It seems amazing  that such a fundamental physical quantity as the
density of cosmic vacuum, comes from a simple combination
$\bar\rho_V = \frac{3 M}{8\pi R_V^3}$ of rather modest
astronomical quantities which are the Local Group mass and the
starting distance of the Hubble flow of expansion.

Thus, the observations of the local expansion flow enable us to
discover local dark energy in the nearby universe and estimate its
density at a distance of a few Mpc from the Milky Way galaxy. The
result is completely independent of the largest-scale cosmological
observations ({\it 1,2}) in which dark energy was first
discovered; it is also compatible with and complementary to them.

Now we discuss the result and its implications.

1. As we already mentioned, the dark energy first revealed itself
in the Hubble flow at very large distances. It was found ({\it
1,2}) that the global cosmological expansion was decelerated by
gravity at times earlier than at the redshift $z = z_V \simeq 0.7$
(which corresponds to a distance $\sim 1000$ Mpc) and accelerated
by antigravity at times later than $z = z_V$ . At the redshift $z
= z_V$, the antigravity of dark energy and the gravity of matter
(baryons and dark matter) balance each other for a moment. The
balance condition is $\rho_M (z_V) - 2 \rho_V = 0,$ where $\rho_M
(z)$ is the cosmological matter density. Since the matter density
scales with redshift as $(1 + z)^3$ and the present-day matter
density is known, $\rho_M (z=0) \simeq 0.3 \times 10^{-26} kg/
m^{3}$, the estimate of the global dark energy density comes from
the balance relation: $ \rho_V = \frac{1}{2}\rho_M (z=0) (1 +
z_V)^3$ (see its numerical value in the beginning of the paper).

In our search for the local dark energy, we have followed exactly
the same logic. Indeed, the zero-gravity radius of Eq.3 is an
exact local counterpart of the ``global'' redshift $z_V$: they
both indicate the gravity-antigravity balance. But what is
temporal globally proves to be spatial locally: the balance takes
place only at one proper-time moment (at $z = z_V$) in the
Universe as a whole, while it exists all the time since the
formation of the Local Group at only one distance ($R = R_V$) from
the group center. Unfortunately, the accuracy of the determination
of $R_V$ is still considerably lower than in the case of $z_V$;
this is mainly because of a relatively small  number of galaxies
-- only two dozens -- in the observed local flow.

The global studies ({\it 1,2}) are reasonably treated as direct
probe of dark energy -- contrary, for instance, to implications
from CMB studies ({\it 3,4}) which are considered indirect. In the
same sense, our local method is the direct one.

2. Our model  leads to an important specific prediction. It
follows from Eqs.1-3 that at distances $R > R_V$, the velocities
of the local expansion flow must be not less than a minimal
velocity $V_{esc}$. The minimal velocity comes from the minimal
total mechanical energy needed for a particle to escape from the
gravitational potential well of the Local Group. Actually, this
prediction may serve as a critical test for the model.

In Fig.1, the minimal velocity $V_{esc}$ is shown by a bold curve;
it turns to zero at $R = R_V$ and grows nearly linearly at $R >
R_V$. This is one curve of a bunch of the curves that cross the
distance segment from 2.1 to 2.3 Mpc corresponding to the observed
position of the galaxy I5152 on the diagram. At $R > R_V$, the
bunch leaves all the 20 other galaxies above the critical curves.
The bunch parameters are the mass of the Local Group $M$ and the
dark energy density $\bar \rho_V$, and if the mass is taken to be
$M = 1.3 \pm 0.3 \times 10^{12} M_{\odot}$ (see above), then the
local dark matter density must be
% 6
\be \bar\rho_V = (0.6 \pm 0.3) \times 10^{-26} \;\; kg/m^3. \ee
\noindent  Thus, the model passes the test with these parameters,
and in this way, the diagram of Fig.1 leads to a new independent
estimate of the dark energy density. The value of Eq.6 is
compatible with the interval of Eq.5.

As is seen in Fig.1, the velocity-distance structure of the flow
follows the trend of the minimal velocity: the linear regression
line of the flow (the thin line) is nearly parallel to the minimal
velocity curve, at $R > R_V$.

A stronger condition may also be checked which requires that all
the 21 galaxies at $R > R_V$ (including the galaxy I5152) are
above the critical lines. In this case, the value of Eq.6 gives an
upper limit for the local dark energy density.

Note that the test is rather sensible: for instance, with a higher
value of the local dark energy density, say, $1.5 \times 10^{-26}$
kg/m$^3$, over half of the galaxies would lie below the curve of
the minimal velocity.

For a comparison, a similar minimal escape velocity is shown also
for a ``no-vacuum model'' with zero dark energy density -- dashed
line in Fig.1. The real flow ignores obviously the trend of the
minimal velocity in this case: the velocities of the flow grow
with distance, while the minimal velocity decreases. It is seen
also that two galaxies of the flow violate obviously the no-vacuum
model: they are located below the dashed line. This comparison is
clearly in favor of the vacuum energy model and against the model
with no dark energy.

3. Another independent test of the model involves the measured
value of the local expansion rate $H = 72 \pm 6$ km/s/Mpc ({\it
11}). Indeed, the model predicts that the expansion rate must be
near the value of $H_V$ (see Eq.4), at distances larger than, say,
2 Mpc. So putting roughly $H = H_V$, we get from this equality a
new estimate for the local dark energy density:
% 7
\be \bar \rho_V = {\frac{3}{8 \pi G}} H^2 = (1 \pm 0.2) \times
10^{-26} kg/m^3. \ee The result is compatible with Eqs.5,6, hence
the model passes this test as well.

Interesting enough, the three seemingly unrelated quantities --
the Local Group mass $M$, the starting distance of the expansion
flow $R_V$ and the expansion rate $H$ -- prove to be essentially
linked, so that $H^2 R_V^3/(G M) \sim 1$. In this fact, the
self-consistency of the model manifests itself.

4. According to recent studies by Sandage and his colleagues (see
a summarizing paper ({\it 17}) and references therein), a regular
Hubble flow of expansion is observed over a very large distance
range from 4 to 200 Mpc. The flow exhibits the Hubble
velocity-distance law, and its expansion rate $H_0 = 62.3 \pm 6.3$
km/s/Mpc is practically the same over the whole scale range. The
simple model of Eqs.1-5 cannot be applied in this case directly.
But our dynamics analysis above (see also papers ({\it 18-21}))
suggests that the kinematic regularity of the flow is possible
only due to the smoothing effect of the perfectly uniform dark
energy on the otherwise lumpy gravitational force field of the
chaotic and non-uniform distribution of the galaxies. In this
case, the rate of expansion must be near the universal value $H_V$
of Eq.4.

With this new understanding, the data ({\it 17}) may be used to
estimate the local dark energy density on the scales 4-200 Mpc.
Using the equality $H_0 = H_V$, we have:
% 8
\be \bar \rho_V = {\frac{3}{8 \pi G}} H_0^2 \simeq (0.74 \pm 0.2)
\times 10^{-26} kg/m^3. \ee This value is practically equal to the
global dark energy density $\rho_V$.

5. Beyond the Local Group's neighboring expanding population which
we examined here, small galaxy groups have long been known to be
quite common; recent studies demonstrate this definitely ({\it
22,23}). Computer identified groups from observational galaxy
catalogs ({\it 24}) have been shown to have an expanding
population via a Doppler shift number asymmetry relative to the
brightest member. Large N-body $\Lambda$CDM cosmological
simulations ({\it 25-28}) show that a structure with a massive
group in its center and a cool expansion outflow outside is rather
typical for scales of a few Mpc and more. The relative numbers of
simulated groups of different kinds ({\it 29}) are near the
observed ones, if the local dark energy density is assumed at the
level of Eq.8. Such studies of other galaxy groups complement
usefully our approach to the dark energy detection around the
Local Group.

\section*{References}

\noindent 1. Riess A.G., Filippenko A.V., Challis P. et al. AJ,
116, 1009 (1998)

\noindent 2. Perlmuter S., Aldering G., Goldhaber G. et al. ApJ,
517, 565 (1999)

\noindent 3. Spergel D.N. et al.  ApJS 148, 175 (2003)

\noindent 4. Spergel D.N. et al.  astro-ph/0603449 (2006)

\noindent 5. Gliner E.B.  Sov.Phys. JETP 22, 378 (1966)

\noindent 6. Karachentsev I.D., Sharina M.E., Makarov D.I., et al.
A\&A, 389, 812 (2002)

\noindent 7. Karachentsev I.D., Makarov D.I., Sharina M.E., et al.
A\&A, 398, 479 (2003)

\noindent 8. Karachentsev I.D., Kashibadze O.G.  Astrofizika 49, 5
(2006)

\noindent 9. Karachentsev I.D., Tully B., Dolphin A.E., et al. AJ
133, 504 (2007)

\noindent 10. Karachentsev I.D. AJ 129, 178 (2005)

\noindent 11. Karachentsev I.D. , Dolphin A.E., Tully, R.B. AJ
131, 1361 (2006)

\noindent 12. Karachentsev I.D., Karachentseva V.E., Huchtmeier
W.K., Makarov D.I., 2004, AJ, 127, 2031

\noindent 13. Chernin A.D., Karachentsev I.D., Valtonen M.J.et al.
 A\&A 415, 19 (2004)

\noindent 14. Chernin A.D., Karachentsev I.D., Valtonen M.J. et
al.  A\&A (2007 -- in press)

\noindent 15. Chernin A.D., Teerikorpi P., Baryshev Yu.V. A\& A
456, 13 (2006)

\noindent 16. Byrd G.G., Valtonen M.J., McCall M., Innanen K.
 AJ 107, 2055 (1994)

\noindent 17. Sandage, A., Tamman, G.A., Saha, A., et al. ApJ 653,
843 (2006)

\noindent 18.  Chernin A.D., Teerikorpi P., Baryshev Yu.V.
(astro-ph//0012021) = Adv. Space Res. 31, 459 (2003)

\noindent 19. Karachentsev, I.D., Chernin, A.D., Teerikorpi, P.
Astrofizika 46, 491 (2003)

\noindent 20. Teerikorpi, P. Chernin, A., Baryshev, Yu., , A\&A
440, 791 (2005)

\noindent 21. Thim, F., Tammann, G., Saha, A., et al.  ApJ, 590,
256 (2003)

\noindent 22. van den Bergh, S.  AJ 124, 782 (2002)

\noindent 23. van den Bergh, S.  ApJ 559, L113 (2001)

\noindent 24.  Valtonen, M. J. and Byrd, G. G.  ApJ  303, 523
(1986)

\noindent 25. Nagamine, K., Cen, R., Ostriker, J. P.  Bul. Amer.
Astron. Soc. 31, 1393 (1999)

\noindent 26. Ostriker, J. P., Suto, Y.  ApJ 348, 378 (1990)

\noindent 27.  Strauss, M. A., Cen, R., Ostriker, J.P.  ApJ 408,
389 (1993)

\noindent 28. Macci\`{o}, A.V., Governato, F. Horellou, C. MNRAS
359, 941 (2005)

\noindent 29. Niemi, S.-M. et al. (to be published)

\noindent 30. A.C., Yu.E., V.D. and L.D. were partly supported by
a RFBR grant 06-02-16366.

\section*{Figure caption}

Fig.1. The Hubble diagram for the very local (distance $R < 3$
Mpc) universe based on {\em A catalog of Neighboring Galaxies}
({\it 12}). The galaxies of the Local Group are located within the
area of 1 Mpc across. The flow of expansion starts in the
outskirts of the group and reveals the linear velocity-distance
relation (the Hubble law) at $R \ge 2$ Mpc (see also the text).

\end{document}